**Earthquake detection capacity of the Dense Oceanfloor Network system for Earthquakes and Tsunamis (DONET)**

**K. Z. Nanjo [1,2,3,4]\*, Y. Yamamoto[4], K. Ariyoshi[4], H. Horikawa[4], S. Yada[4], N. Takahashi[5,4]**

[1]  Global Center for Asian and Regional Research, University of Shizuoka, 3-6-1, Takajo, Aoi-ku, Shizuoka, 420-0839, Japan

[2]  Center for Integrated Research and Education of Natural Hazards, Shizuoka University, 836, Oya, Suruga-ku, Shizuoka, 422-8529, Japan

[3] The Institute of Statistical Mathematics, 10-3, Midori-cho, Tachikawa, Tokyo, 190-8562, Japan

[4] Yokohama Institute for Earth Sciences, Japan Agency for Marine-Earth Science and Technology, 3173-25, Showa-machi, Kanazawa-ku, Yokohama, Kanagawa, 236-0001, Japan

[5] National Research Institute for Earth Science and Disaster Resilience, 3-1, Tennodai, Tsukuba, Ibaraki, 305-0006, Japan

**\*** Correspondence: nanjo@u-shizuoka-ken.ac.jp; Tel.: +81-54-245-5600

ORCID ID

https://orcid.org/0000-0003-2867-9185 (K. Z. Nanjo)

https://orcid.org/0000-0002-0224-8603 (Y. Yamamoto)

https://orcid.org/0000-0002-6353-1082 (K. Ariyoshi)

https://orcid.org/0000-0003-3914-1348 (N. Takahashi)

Email

nanjo@u-shizuoka-ken.ac.jp (K. Z. Nanjo)

yamamotoy@jamstec.go.jp (Y. Yamamoto)

ariyoshi@jamstec.go.jp (K. Ariyoshi)

horikawah@jamstec.go.jp (H. Horikawa)

syada@jamstec.go.jp (S. Yada)

narumi@bosai.go.jp (N. Takahashi)

**Abstract** We studied the earthquake detection capacity of DONET (Dense Oceanfloor Network system for Earthquakes and Tsunamis) operating in the Nankai Trough, a target region monitored for future megathrust earthquakes. The focus of this paper was to evaluate the impact on this capacity from the malfunction of parts of the network. For this purpose, the completeness magnitude, above which all earthquakes are considered to be detected by a seismic network, was used. Then, a catalog that includes events observed by DONET was used. We found spatiotemporal variability of completeness magnitude, ranging from values below 1 in one of the areas where stations are densely deployed to values above 2 at the periphery and outside of the DONET area. We conducted a simulation computation for cases of malfunction of densely distributed stations. The results showed that completeness estimates in the area near the malfunctioning stations were about 1 magnitude larger. This implies that malfunction repair and/or replacement with new stations would be desirable because they pronouncedly affect earthquake monitoring. We then demonstrated an example of how to use the information of completeness magnitude as prior knowledge to compute the $b$ value of the Gutenberg-Richter distribution. The result indicates the $b$ value as a proxy that can help to image stress heterogeneity when there is a magnitude-6 class slow slip event on the Nankai Trough plate boundary.

**Keywords** Computational seismology, Japan, Nankai Trough, Probability distribution, Seismicity and tectonics, Statistical seismology.

**Statements and Declarations**

**Competing interests** The authors declare no competing interests.

**Ethics approval** Not applicable.

**Consent to participate** All authors agree to participate.

**Consent for publication** All authors agree for publication.

**Conflict of interest** The authors declare no competing interests.

**Article Highlights**

- Earthquake detection capacity of the dense oceanfloor network operating in the Nankai Trough was evaluated.

- Results demonstrate that malfunction of parts of the network pronouncedly impacts detection capacity.

- Information on detection capacity is used as prior knowledge to help to image stress changes due to slow slip events

# 1 Introduction

About 80 years have passed since the last series of Nankai Trough earthquakes occurred, the 1944 Tonankai and the 1946 Nankai earthquakes, both belonging to a magnitude ($M$) 8 class. The Nankai Trough earthquakes occurred with a return period of about 100-200 years. The possibility of large earthquakes in the Nankai Trough has increased. The probability of the occurrence of an impending large earthquake along the Nankai Trough is 70-80% in the next 30 years (Headquarters for Earthquake Research Promotion 2023).

In the wake of the 2004 Sumatra earthquake (e.g., Wiseman and Bürgmann 2011), the Japanese government established a seafloor network of cable-linked observatories around the Nankai Trough (e.g., Ariyoshi et al., 2021a). This network is known as the Dense Oceanfloor Network system for Earthquakes and Tsunamis (DONET: Kaneda et al. 2015; Kawaguchi et al. 2015), and is in operation to constantly monitor earthquakes (Nakamura et al. 2015; Nakano et al. 2016, 2018a) and tsunamis (Maeda et al. 2015) (Figs. 1 and 2). Each station within DONET is equipped with strong-motion seismometers, broadband velocity seismometers, quartz pressure gauges, and differential pressure gauges to detect all types of seafloor movements, from slow movements such as crustal deformation to fast movements such as ground motion generated by earthquakes. DONET was developed and installed by the Japan Agency for Marine-Earth Science and Technology (JAMSTEC), and is currently being transferred to and operated by the National Research Institute for Earth Science and Disaster Resilience (NIED) (Aoi et al. 2020).

JAMSTEC, following the installation of DONET, processed waveform data to create an earthquake catalog for detailed understanding of seismicity along the Nankai Trough (Nakano et al. 2018b; Yamamoto et al. 2022a,b). These are invaluable resources for seismicity-related studies. It is thus vital to establish clear quality benchmarks for the catalog. A common benchmark is the magnitude of completeness, above which all events are assumed to be detected by the seismic network (Rydelek and Sack 1989; Gomberg 1991; Wiemer and Wyss 2002), and this magnitude of completeness was widely used (Kvaerna et al. 2002a, 2002b; Enescu et al. 2007, 2009; Wiemer 2001; Cao and Gao 2002; Wiemer and Wyss 2000; Marsan 2003; Woessner and Wiemer 2005; Amorèse 2007; Tinti and Mulargia 1985; Nanjo et al. 2010a). Such quantification of completeness is a necessary input for virtually any study involving the statistical properties of earthquake populations, for example, rate estimates or estimates of the $b$-value of the Gutenberg-Richter (GR) distribution (Gutenberg and Richter 1994; Ishimoto and Iida 1939), defined in the subsection 3.4. They provide descriptions of completeness for direct use by end users, such as limiting the catalog to a level of completeness that contains only events with magnitudes that are considered to be completely recorded.

We examined earthquake detection probabilities and completeness levels for the catalog created and maintained by JAMSTEC (Nanjo et al. 2023a,b). Among the various existing methods to compute these detection probabilities and completeness levels, we opted to employ the Probability-based Magnitude of Completeness (PMC) method (Schorlemmer and Woessner 2008; Schorlemmer et al. 2010, 2018; Nanjo et al. 2010b; Gentili et al. 2011; El-Hussain et al. 2020; Nanjo 2020a). Using this method, for 52 stations (Fig. 2), detection capabilities were derived over time from empirical data only, namely earthquake information, phase data, station information, and network-specific attenuation relations. From the entire phase-data history, we estimated the operational times of each station, then synthesized detection-probability maps for specific magnitudes and completeness maps. These maps were computed for any time in 2015-2019, a period in which a dataset was available for our study.

Our focus was to examine a possible use of the PMC method, which is used to assess a network's performance (Schorlemmer and Woessner, 2008; Schorlemmer et al. 2010; Nanjo et al. 2010b, Nanjo 2020a). We explored it for cases where we removed stations from DONET to see what effect this had on the magnitude of completeness. We propose the use of this application to investigate the effect of network failures on completeness.

It is of interest to demonstrate the direct use of the completeness obtained in this study as foresight information for seismicity-related studies. We studied the $b$-values computed by using only complete data to investigate whether they were correlated to a slow slip event (SSE) detected by the oceanfloor geodetic system, which functions independently from DONET.

## 2 Data

### 2.1 DONET, individual stations, and SSE

The paper used earthquake catalog data based on the DONET cabled network offshore Japan and focuses on the earthquake detection capability of the network. The method used in this study partly follows the methods of Schorlemmer and Woessner (2008) using statistics on the earthquake catalog and picks to investigate maps of the magnitude of completeness. The present study used estimates of $b$ values as a function of time and compared transients of $b$ values to an SSE event.

In this section, we first explain DONET and recorded earthquakes. We then describe detailed information on individual stations. Finally, we present the SSE that occurred along the Nankai Trough in 2017-2018.

### 2.2 DONET and earthquakes

DONET, one of the densest oceanfloor networks in Japan (Figs. 1 and 2), was constructed to achieve a real-time monitoring system in the central region of the Nankai Trough (Kaneda et al. 2015; Kawaguchi et al.

2015). Its primary objectives are the early detection of earthquakes and tsunamis. DONET1 and DONET2, subnetworks of DONET, cover the areas from which ruptures of the 1944 Showa Tonankai earthquake and the 1946 Showa Nankai earthquake extended to 120 [km] east and 120 [km] west, respectively (Kanamori, 1972). The previous Nankai Trough event, a set of the 1854 Ansei Tokai earthquake and Ansei Nankai earthquake 32 hours later (Ando 1975) had a rupture pattern similar to the 1944/1946 Tonankai/Nankai earthquakes. Thus, the DONET area should be intensively monitored for future mega earthquakes.

DONET consists of 51 stations in total (Fig. 2) and each station is equipped with a ground motion sensing system as well as a pressure sensing system. The former ground motion sensing system consists of a strong-motion seismometer and a broadband velocity seismometer (Nakamura et al. 2015; Nakano et al. 2016, 2018; Maeda et al. 2015), while the latter pressure sensing system consists of a quartz pressure gauge and a differential pressure gauge (Araki et al. 2017; Ariyoshi et al. 2021a,b). Burying the ground motion sensing system into the sediment layer reduced the background noise and other environmental effects (Araki et al. 2008; Kaneko et al. 2009). For DONET1 and DONET2, the looped backbone cable that provides the power feed and communications channel to stations is separately connected among science nodes (Fig. 2c). The node is a device that functions as a hub that connects the stations. DONET1 has five science nodes connecting 22 stations while DONET2 has seven science nodes connecting 29 stations, where each node connects 4 or 5 stations at a station interval of 10-20 [km] and stations are aggregated around each node. All DONET data are transmitted to JAMSTEC, JMA, and NIED in real-time where they are stored. Additionally, one station called KMDB1 (Kopf et al. 2011), which was installed in a borehole, and with a broadband velocity seismometer, was operating during the period of investigation (2015-2019), so we included this station in our analysis (Fig. 2).

As of March 31, 2019 (the most recent time of earthquakes stored in the catalog we used), DONET was operating 41 stations and one borehole station (KMDB1), as shown in Fig. 2a,b. The operation of Node A (4 stations) and Node E (5 stations) of DONET1 was interrupted since 2018 and 2016, respectively, due to a defective connection between these nodes and the backbone cable caused by a malfunction of these nodes. The operation of station MRF22 connecting Node F of DONET2 was also interrupted since 2018 due to a communication failure between the node and the station.

JAMSTEC is recording seismicity in and around the central Nankai Trough with DONET, including ordinary and very low-frequency earthquakes, and tremors. JAMSTEC created an earthquake catalog that includes local seismicity of ordinary earthquakes between Oct. 2015 and Mar. 2019. It was created by using waveform data of broadband velocity seismometers, while strong-motion seismometers are rarely used. Algorithm that is divided into event detection and location, both conducted by JAMSTEC, is briefly explained as follows: The method of Horiuchi et al. (2010) was used for event detection and automatic arrival time picking. Operators visually revised the results of automatic picking for events that were detected by three or more DONET stations, where we used only P-phases recorded and picked by DONET stations because of the greater accuracy of P-phase detection relative to S-wave detection. Note that there many of events regarded as detected events with the automatic arrival time picking are discarded by operators' visual inspection. In the operation associated with the visual revision, operators additionally checked waveform data for all DONET stations to conduct manual picking. The event location would be possible to be carried out by using data from only DONET stations, but to improve the reliability of them, JAMSTEC retrieves data from onshore stations. Detected events are likely possible to be located together with estimates of their magnitudes, if these events are located in and around the authoritative region of DONET (this region is defined in the next paragraph, and seen in Fig. 1). So, our interest in the present study was more in the ability to detect earthquakes rather than to locate earthquakes. Note that the onshore stations that are not connected to the detecting algorithm were not included in our PMC analysis (we considered only DONET stations shown in Fig. 2). There is an exception for a specific period (Sep. 2016 to Mar. 2017) in which operators did not check them for all stations (see the subsection 2.3).

To define our study region, we used the authoritative region (134-138°E, 32-34°N) to cover the extent of DONET and a depth range of 0-60 [km]. For this study region, the catalog contains local seismicity of 11,538 ordinary events for $M$=-0.4 to 5.9 from Oct. 2015 to Mar. 2019 (Fig. 1). Our interest was the earthquake detection capacity along the Philippine Sea plate. Seismicity in the Philippine Sea plate in the coupling zone is considered to reflect interplate partial locking between the subducting plate and the overriding plate (Matsumura 1997). Because we wanted to investigate seismicity both around the plate boundary and within the Philippine Sea plate, we wanted to remove the crustal events in the overriding plate. We retrieved from the catalog, any earthquakes in the depth range from 5 [km] above the plate boundary (or the upper surface of the Philippine Sea plate; Nakanishi et al. 2018) to 25 [km] below the boundary. The shallower limit of the depth range, namely a depth of 5 [km] above the plate boundary, was used to ensure that a usual depth error was taken into consideration. Since $P_E$ (earthquake detection probability) and $M_P$ (magnitude of completeness), both defined in the subsection 3.3, can be computed for points in space and time, the depth at which these values are computed needs to be defined. This is referring to three dimensions, so that's why we considered the depth. To find a meaningful depth, we investigated cross-sectional distributions of earthquakes. Visual inspection of the three selected cross-sections (Fig. 1b) shows that a depth of 10 [km] below the plate boundary is an appropriate depth to represent seismicity along the Philippine Sea plate. We defined it as the depth for which we computed $P_E$ and $M_p$.

JAMSTEC routinely computes local magnitude ($M_L$), where for sake of simplicity we used $M$ instead of $M_L$. To compute $M$, the attenuation relation of Watanabe (1971) is used: $0.85M-2.50=\log A_v+1.73\log r$ for hypocentral distance $r<200$ [km] and $0.85M-2.50=\log A_v+1.73\log r+0.0015(r-200)$ for $r\geq 200$ [km], where $A_v$ is the maximum amplitude of the vertical component of velocity. The determination of $M$ is consistent with the range of magnitudes of interest for DONET.

As a basis of the seismicity study, understanding magnitude scales is critical. We examined whether large (small) earthquakes in our catalog were indeed large (small), as in the JMA catalog. Namely, we compared if both magnitude scales are similar. We used earthquakes in the authoritative region (134-138°E, 32-34°N, depth of 0-60 [km]) for both catalogs. Then, an earthquake in the DONET catalog was paired with that in the JMA catalog if the time difference between them was within two seconds and the epicentral distance between them was less than 30 [km], while ignoring one-to-multiple cases. A list of paired earthquakes (Fig. 4a) shows that magnitude of the DONET catalog was positively correlated with magnitude of the JMA catalog ($M_{JMA}$). A similar correlation was observed in a previous study (Yamamoto et al. 2022). We also found that the slope of the least-square regression line between these magnitudes (black line) was less than one, due to the tendency that the magnitude of the DONET catalog, $M$ (local magnitude), is larger than $M_{JMA}$ for $M_{JMA}$ from 0 to 2~3. To show that this tendency was location-dependent, the study region was divided into a nearshore region (33-34°N) and an offshore region (32-33°N) (Fig. 3b,c). The slope of the regression line for the former region (Fig. 3b) was larger than that for the latter one (Fig. 3c). This suggests that for the offshore region, the tendency of $M$>$M_{JMA}$ was more pronounced with decreasing $M_{JMA}$. The relation between $M$ (local magnitude) and $M_{JMA}$ will be debated in the "Discussion" section.

Although sophisticated regressions such as the weighted linear regression (weighted least squares) are available, a simple approach was taken to capture essential aspects of the relation between $M$ and $M_{JMA}$. That is, we took the ordinary linear regression (ordinary least squares) and showed the results using the black line in Fig. 3a-c.

## 2.3 Operational times for individual stations

To compute operational times of all stations for the entire period of investigation, picks recorded by each station to detect an earthquake were plotted as a function of time (green dots in Fig. 2a,b). All stations were considered to have started operation in Oct.-Dec. 2015, except for KMC21, MRD16, and MRD17, whose start of operation was delayed compared with the other stations: KMC21 started operation in Jun. 1, 2016, and MRD16 and MRD17 started operation in Mar. 1, 2016. All stations were operational until the end of the investigation period (Mar. 31, 2019), except for four stations connecting Node A of DONET1, five stations connecting Node E of DONET1, and MRF22. The reason for these exceptions is a malfunction of these nodes and/or a node-station communication failure. Some stations showed one or more off-time periods less than one month. Off-time periods are commonly observed at stations connected to respective nodes. For example, stations connecting Node A of DONET1 exhibited irregular off-time patterns between Sep. 2016 and Mar. 2017. During this period, these stations and Node A were actually operating, and data were transmitted to JAMSTEC. The irregular patterns are attributed to a slight change of internal protocol at JAMSTEC for data processing associated with this node during the corresponding period. Stations connecting Node G of DONET2 also exhibited similar irregular patterns to those of Node A of DONET1, for the same reason as was indicated for Node A. Stations connecting

Node D of DONET1 exhibited two off-time periods starting in Feb. 2016 and Sep. 2018, because of a sequential failure and repair of Node D, respectively.

For each station, it was necessary to define the time range from which pick information was sampled to compute detection-probability distributions (Fig. 3b), as described in the "Methods" section.

Off-time periods for some stations were short and negligible compared with their entire lifetime, so we decided to use the entire lifetime (grey bars in Fig. 2a,b).

It would be beneficial to show how the noise levels varied over times. This is related to the noise level in a specific frequency band for the DONET stations (Nakano et al. 2019). Temporal variations in the running spectra at KMB05 shown in Fig. 1 was examined (Supplementary Figs. S1-S3). Regardless of day or night,

there was no pronounced temporal variation in the running spectra in a frequency band of 2-20 Hz on which

we focused for earthquake detection of DONET (Supplementary Figs. S1a, S2a, and S3a). Seasonal

variation was generally seen in the microseism frequency band (0.1-1 Hz), but not in the high frequency band

of 2-20 Hz (Supplementary Figs. S1b, S2b, and S3b). Comparison between 2018 and 2019 (Supplementary

Figs. S1a and bottom-right panel of Supplementary Figs. S1b) shows that the running spectra in the frequency band of 2-20 Hz of the up-down component remained unchanged after a year in general. The same applies for the east-west and north-south components (Supplementary Figs. S2 and S3). Our observation shows that noise levels generally did not vary over times in the frequency band (2-20 Hz), implying that the temporal distribution of low-magnitude events homogeneous over time.

## 2.4 SSE

As shown in the "Results" section, information on the resultant completeness is applicable to seismicity analysis. As an example, we investigated the relation between the spatiotemporal changes in $b$ and the occurrence of the $M$6.6 SSE (Yokota and Ishikawa 2020). This SSE was detected on the shallow plate boundary near the trench axis in 2017-2018 (Fig. 1) based on seafloor geodetic data obtained by the Global Navigation Satellite System-Acoustic (GNSS-A) combination technique. The SSE involved thrust faulting along the plate boundary with slip direction nearly perpendicular to the trench axis. This occurred at the periphery of the DONET2 area. The SSE of $M$6.6 is, to our knowledge, the largest among reported SSEs on the shallow plate boundary near the trench axis. SSEs belonging to an $M$6-7 class have often been observed on the deep plate boundary (Ozawa et al. 2016; Kobayashi 2017). These deep SSEs that were not located in and around the DONET area were out of scope of our study. JAMSTEC (Araki et al. 2017; Ariyoshi et al. 2021a) reported 12 SSEs that occurred during 2011-2020 in the DONET1 area on the shallow plate boundary near the trench axis. However, these belong to an $M$5-class.

Various slow earthquakes, including not only SSEs but also tremors and very low-frequency events, occur along the Nankai Trough (Tamaribuchi et al. 2022; Takemura et al. 2022). The observation of slow earthquakes' analogies and differences has a multidisciplinary value with respect to the physical process of the plate boundary (Yokota and Ishikawa 2020). Our future research aims to conduct comparative studies among various slow earthquakes (SSEs, tremors, and very low-frequency events) and ordinary (fast) earthquakes. As a first attempt, we considered one example in the "Results" section, comparing an SSE with local seismicity of ordinary (fast) earthquakes based on $b$ values.

## 3 Methods

### 3.1 PMC method and GR distribution

In this section, we first explain the PMC method. This method relies on two sources of data: (1) station data describing the location for each station in the network; (2) earthquake catalogs describing the location, time, and $M$ for each earthquake including data describing which stations were used to detect this earthquake, where $M$ is used to express local magnitude ($M_L$) in this study (see also the subsection 2.2). The method is divided into an analysis section and a synthesis section. We recommend the readers to also appreciate a previous application (Schorlemmer and Woessner 2008).

We then briefly introduce the concept of the GR frequency-magnitude distribution (Gutenberg and Richter, 1944; Ishimoto and Iida, 1939). We provide an example of using the result of completeness estimates as prior information to compute the parameters of GR distributions.

## 3.2 PMC method: analysis part

With this method, we wanted to investigate the phase picking capacity of P-phases depending on the hypocentral distance. In the analysis part, data triplets were first compiled for each station, following the approach described in Schorlemmer and Woessner (2008). A triplet contains, for each earthquake, (i) information on whether or not this station was used for detecting the event, (ii) the $M$ of the event, and (iii) its hypocentral distance $L$ from the station. Fig. 4a provides an illustrative example to show how to generate data triplets. The top panel of Fig. 4a shows that the station was used to detect events 1, 2, and 4 (green), but not events 3 and 5 (red), where the $i$-th event is identified by the index in $M$ and $L$ ($M_i$ and $L_i$) ($i = 1, 2, \ldots,$ 5). Note that events 3 and 5 were recorded by using other stations. The data triplet of event 1 contains (i) the fact that the station was used to detect this event (green), (ii) the magnitude ($M_1$), and (iii) its distance from the station ($L_1$). If a station was used to pick an event, the data triplet of this event was referred to as the "plus triplet", otherwise as the "minus triplet". Data triplets of events 1, 2, and 4 are plus triplets, and those of events 3 and 5 are minus triplets. These triplets were plotted in a graph of $L$ as a function of $M$. The same applies for all stations.

The illustrative example (Fig. 4a) is based on the premise of picking an event with small $M$. If an event with large $M$ occurs, the phase picking algorithm -depending on the network operators' protocol- may abort to avoid operator overloading. Namely, there is a bias that all of the triplets are minus ones for $L$ larger than a threshold distance and for $M$ larger than a threshold magnitude, both predefined by network operators. For the ranges of $M$ and $L$ ($0 \leq M \leq 3$ and $0 \leq L \leq 200$ [km] in Fig. 4b) considered in this study, there is no such bias.

Using triplets for each station, the desire was to determine $P_D(M, L)$, detection probability for events of magnitude $M$ at hypocentral distance $L$ at a station (This is identical with $P_D(M, L)$ defined in Eq. 9 of Schorlemer and Woessner 2008). Using data triplets close to a given pair $(M, L)$, $P_D(M, L)$ was computed based on the number of plus triplets, $N_+$ (green plus symbol), divided by the sum of $N_+$ and the number of minus triplets, $N_-$ (red minus symbol): $P_D(M, L) = N_+/(N_+ + N_-)$ (Fig. 4b).

Triplets were selected by measuring the distance between each triplet and a given pair $(M, L)$. To measure such a distance, a metric in the $M$-$L$ space needs to be defined. Schorlemmer and Woessner (2008) proposed the use of an attenuation equation for earthquakes located in a given local seismic network, and redefined a metric in the transformed magnitude-magnitude space ($L_M$). In this study, we followed this idea (Eq. 8 of Schorlemmer and Woessner 2008) and used the attenuation equation of Watanabe (1971). We selected all triplets that obey the criterion $L_M \leq 0.4$. The value of 0.4 is a usual magnitude error (roughly the mean of the standard deviations of magnitudes determined by JAMSTEC).

$P_D(M, L)$ was smoothed by applying a simple constraint: $P_D$ cannot decrease with a smaller $L$ for the same $M$. This smoothing accounts for high probabilities at short distances (Fig. 4b). Another constraint was also applied, namely that the smoothed probability cannot increase with decreasing $M$ at the same distance.

This approach assumes that single earthquakes occur at different times because multiple earthquakes that had occurred at different locations at the same time were not reported by JAMSTEC during 2015-2019.

Fig. 4b shows the distribution of data triplets of KMB05 connecting Node B of DONET1 (top panel). To create this panel, we used triplets during the period Nov. 1, 2015-Mar. 31, 2019, which is shown by the grey bar for KMB05 in Fig. 2a. Based on this distribution of triplets, we created the bottom panel showing the corresponding distribution of $P_D(M, L)$ to which smoothing constraints were applied. The middle panel is the same as the bottom one for the case in which the smoothing constraints were not applied. This panel is an intermediate product to show the effect of the smoothing constraints on $P_D$ (bottom panel). A non-smoothed distribution (middle panel) was not used for further analyses.

There was no triplet for short hypocentral distances ($L<10$ [km]) due to the distances from stations to earthquakes along the subducting Philippine Sea plate. To create Fig. 4b, earthquakes with depths equal to or shallower than a depth of 5 [km] above the plate boundary were not used because we wanted to remove the crustal events in the overriding plate. $P_D>0.8$ (green) in $M$=0.5~2.5 was found for $L$=0~30 [km], highlighting that the dense configuration of stations connecting Node B enhanced the detection capacity of earthquakes for such values of $M$ and $L$. Detection probabilities decreased with increasing distances while keeping the magnitude fixed. This is in accordance with the expectation that the probabilities should be higher for smaller distances and higher magnitudes.

3.3 PMC method: synthesis part

In the synthesis part, basic combinatorics were used to obtain $P_E(M, x)$, the detection probabilities for earthquakes of $M$ at location $x$ (This is identical with $P_E$ defined in Schorlemmer and Woessner 2008). $P_E(M, x)$ for DONET is defined as the probability that three or more stations detect an earthquake of $M$ at $x$. The minimum number of stations must be adjusted if the DONET condition is based on another number of stations.

If $P_E$ is larger than the threshold, this is likely that an earthquake will not be missed. Previous researchers arbitrarily selected several values below but close to 1 as their threshold (Schorlemmer and Woessner 2008; Nanjo et al 2010b; Schorlemmer et al. 2010, 2018; Gentili et al. 2011; El-Hussain et al. 2020; Nanjo 2020a). Given the considerable uncertainty of all phenomena such as tectonic strain/stress accumulation, earthquakes, and others in this study, the threshold needs to be as high as possible to ensure that an earthquake will not be missed. The magnitude of completeness $M_p(x)$ is given as $M$ above which earthquakes are detected with $P_E(M, x) \geq 1-Q$, where $Q$ is the complementary probability that events will be missed (this is identical with $P_E$ defined in Schorlemmer and Woessner 2008). The choice of the $Q$ value is arbitrary and should reflect desired accuracy. We assumed $Q=10^{-6}$, a value lower than the values adopted by previous studies, because we desired the most conservative value.

## 3.4 GR distribution

GR distribution is given as $\mathrm{Log}N=a-bM$, where $N$ is the number of earthquakes with magnitudes larger than or equal to $M$, while $a$ and $b$ are constants. $a$ characterizes seismic activity or earthquake productivity of a region and $b$ is used to describe the relative occurrence of large and small events (i.e., a high $b$ value indicates a larger proportion of small earthquakes, and vice versa). Spatial and temporal changes in $b$ are known to reflect the stress state of the Earth's crust (e.g., Smith 1981; Schorlemmer et al. 2005; Narteau et al.

2009) and are influenced by asperities and frictional properties (Hirose et al. 2002; Yabe 2003; Schorlemmer and Wiemer 2005; Nanjo 2020b), and by an interface locking along subduction zones (Sobiesiak et al. 2007; Ghosh et al. 2008; Nanjo and Yoshida 2018). In the laboratory, $b$ values are inversely dependent on differential stress (Scholz 1968). If this dependence holds in the Earth's crust (Scholz 2015), then measurements of spatial and temporal changes in $b$ could act as a 'stress meter'. To estimate $b$ values consistently over time and space, information on completeness magnitudes is needed. Below the state of completeness, a fraction of events is missed by the observation network because: (1) they are too small to be recorded by enough stations; (2) network operators decided that events below a certain threshold are not of interest; (3) in the case of an aftershock sequence, because they are too small to be detected within the coda of larger events (Woessner and Wiemer 2005). Thus, an earthquake sample used to compute the $b$ value must be above or equal to the level of completeness. Incorrectly estimated completeness may subsequently lead to wrong results when computing $b$ values, and underestimation of the completeness level would lead to an underestimation of the $b$ value (Schorlemmer and Woessner 2008).

Among existing methods to compute $b$ values, the most basic and computationally inexpensive method is the Maximum Curvature (MAXC) method (Wiemer and Wyss 2000; Wiemer 2001). The MAXC method is applied to an earthquake sample to compute its own completeness magnitude, $M_{c(MAXC)}$. The obtained value for $M_{c(MAXC)}$ is known to be an underestimate (Woessner and Wiemer 2005). For computing $b$, the maximum-likelihood method is applied to earthquakes with magnitudes $\geq M_{c(MAXC)}+M_{corr}$, where $M_{corr}$ is a correction value (Woessner and Wiemer 2005). One challenge is how to select a value for $M_{corr}$. Although $M_{corr}=0.2$ was recommended by Woessner and Wiemer (2005) for California, it remains uncertain whether this can be considered ubiquitous for global locations. To define $M_{corr}$ or $M_c+M_{corr}$, we used the information on $M_p$, as shown in the "Results" section. $M_{corr}$=0.2 is not applicable but $M_{corr}$=0.5 seems to be more appropriate for our case.

We used the test proposed by Utsu (1992, 1999) to examine whether the difference in $b$ is significant. If $\log P_b$, the logarithm of the probability that the $b$ values are not different, is equal to or smaller than -1.3 ($\log P_b \leq$-1.3), then the difference in $b$ is significant (Schorlemmer et al. 2004; Nanjo and Yoshida 2017; Nanjo et al. 2019; Nanjo 2020b). We also introduced bootstrapping errors (Schorlemmer et al. 2003; Nanjo and Yoshida 2017; Nanjo et al. 2019; Nanjo 2020b) to show an error bar (one standard deviation) of $b$ and $M_c$+$M_{corr}$.

## 4 Results

### 4.1 Mapping $P_E$ and $M_p$ in and around DONET

Fig. 5a-c shows the spatial distribution of $P_E$ for $M$=1.0, 1.5, and 2.0 on Jan. 1, 2019. In creating the map of $P_E$($M$=1.0, $x$) in Fig. 5a, $x$ is a grid point of a 0.05° × 0.05° grid on the curved surface located at a depth of 10 [km] below the upper surface of the subducting Philippine Sea plate (Nakanishi et al. 2018). $P_E$($M$=1.0, $x$) for $x$ is based on a set of 42 $P_D$($M$=1.0, $L$) values for 42 stations operating on Jan. 1, 2019 (Fig. 2a,b), where $L$ is the distance from $x$ to a station. The distribution of $P_D$($M$, $L$) for each station was computed based on data triplets for the period shown in a grey bar in Fig. 2a,b. Thus, because $P_E$($M$=1.0, $x$) can be computed at any time point in the investigation period (Oct. 1, 2015-Mar. 31, 2019), we selected a specific time point, namely Jan. 1, 2019, and then computed $P_E$($M$=1.0, $x$) to create Fig. 5a. The reader may note that time windows such as the one from Jul. 1, 2018 to Jan. 1, 2019 were not assumed to compute $P_D$($M$=1.0, $L$) and that the map of $P_E$($M$=1.0, $x$) in Fig. 5a is independent on the volume where earthquakes were selected from. Thus, this map was created for a depth of exact 10 [km] below the upper surface of the subducting Philippine Sea plate. The same applies to $M$=1.5 and 2.0. As expected, the three maps (Fig. 5a-c) show that $P_E$ increased with $M$. We plotted, in Fig. 5a-c, earthquakes in the depth range from 5 [km] above the plate boundary to 25 [km] below the boundary during Oct. 16, 2018-Mar. 31, 2019, where station configuration was unchanged (Fig. 2a,b), including our selected time point, Jan. 1, 2019. These earthquakes were located in areas with predominantly high probabilities, supporting consistency between our synthesis and the observation. Fig. 5a shows spatial variability of detection probabilities for $M$=1.0: a large $P_E$ is highlighted in areas near the stations. In particular, $P_E \geq 0.9$ (green) for $M$=1.0 was observed in three areas: the DONET1 area, the area near Node A, B, …, and F of DONET2, and the area near Node G of DONET2. Although $P_E$ for $M$=1.5 was generally > 0.9 over almost the entire DONET area (Fig. 5b), we again observed a similar variability and strong decrease towards the outside of the DONET area. $P_E \geq 0.9$ for $M$=2.0 (Fig. 5c) was observed over almost the entire study region (i.e., the authoritative region of DONET).

We found that $M_p$ for Jan. 1, 2019 (Fig. 6a) varied from values <1.0 (yellow to orange) to values >2.0 (red to blue) in and around DONET. $M_p$ values were generally <2.0 in areas of DONET1 and DONET2 while $M_p$ values were >2.0 in areas between them. This is due to the absence of stations in between. Low $M_p$ values (<1.5, orange to red) are associated with three areas near Nodes B-D of DONET1, and Nodes B-E and G of DONET2. In areas near Nodes D and E of DONET2, $M_p$ values were <1.0 (yellow to orange). As expected, a strong increase towards the outside of the DONET area was observed. In and around the source area of the $M$6.6 SSE, the maximum of $M_p$ values (max $M_p$) were ~2.

To demonstrate how a change in network configuration influences the capacity to detect earthquakes, we show the spatiotemporal evolution of $M_p$ in Fig. 6a-c, indicating snapshots on Jan. 1 for 2019, 2018, and 2016. Similar to $P_E$, $M_p$ can be computed at any time point in the investigation period. We adopted three specific time points, as noted above. We did not show the snapshot for 2017 because it was the same as that for 2018 (i.e., there was no difference in network configuration between 2018 and 2017), as shown in Fig. 2a,b. The detection level changed as operational stations changed. The failure of Node E of DONET1 in 2016 (Fig. 6b,c) and of Node A of DONET1 in 2018 (Fig. 6a,b) pronouncedly influenced $M_p$, resulting in a better detection capacity in the northern area of DONET1 in 2016 than in 2019 (Fig. 6a,c). Other detailed temporal changes in $M_p$ were associated with terminating or starting operations of individual stations, as shown in Fig. 6a-c. The three snapshots show that max $M_p$~2 in and around the source area of the $M$6.6 SSE did not pronouncedly change, irrespective of time.

## 4.2 Virtual removal of one or more stations from DONET

The previous section implies a remarkable effect of major system failures such as node malfunction on the detection capacity of the network. To investigate this effect, scenario computations were performed by virtually removing existing stations from the most recent network configuration (Jan. 1, 2019 in Fig. 6a). We compared the probability-based completeness magnitudes between the virtual case, $M_{pv}$, and the real case, $M_p$, for all locations in the study region with a grid spacing of 0.05° × 0.05°: $\Delta M_p = M_{pv} - M_p$. Fig. 7a shows the map of $M_{pv}$ for virtual station removal of Node E of DONET2 from the existing network, where the mapping procedure was the same as for Fig. 6a, but $P_D$ was not used for stations of Node E of DONET2. The most likely scenario is a malfunction of this node. When excluding the Node E stations of DONET2, completeness magnitudes were larger than or equal to the completeness magnitude of the stations. In detail, the $\Delta M_p$ map in Fig. 7c shows that completeness estimates, excluding Node E stations of DONET2 in the area near this node, were about 1 magnitude larger (orange). This map also shows a general decreasing trend of $\Delta M_p$ with increasing distance from this node. Removing another Node D next to the removed Node E, which would imply a further enhanced scenario of node malfunction (Fig.

7b), was considered next. In detail, completeness estimates, excluding stations of Nodes D and E of DONET2 in the area near these nodes, were about 1.2 magnitudes larger (red) (Fig. 7d). Similar to Fig. 7c, there was a general decreasing trend of $\Delta M_p$ with increasing distance from Nodes D and E of DONET2 (Fig. 7d). The spatial extent of the influence was broader.

The area that covers the extent of Nodes D and E is located near the region which straddles the rupture zone of the 1944 Showa Tonankai earthquake and the rupture zone of the 1946 Showa Nankai earthquake. Near this region, the former rupture started to propagate east while the latter one propagated west. Future Nankai Trough earthquakes may have a similar rupture characteristic to the 1944/1946 Tonankai/Nankai earthquakes. Given that the sites near Nodes D and E of DONET2 in Fig. 6a showed the lowest $M_p$ values (<1), failure of Nodes D and E affecting earthquake monitoring would be the most pronounced among these nodes' failures. Repairing these nodes or replacing them with new ones when their malfunction occurs would be desirable.

On the other hand, our scenario computation resulted in $\Delta M_p$<0.4 (white in Fig. 7b,d) at almost all sites except for the area that covers the extent of Nodes D and E, where the usual magnitude of error of DONET was 0.4, as described in the "Data" section. This can be interpreted as an indication that the success of aggregated configuration of stations around the respective nodes enhanced the stability of event detection capacity.

4.3 Applied example by using the result of $M_p$

Many papers that employed a completeness study argued that for virtually any statistical analysis of earthquake catalogs, the knowledge of completeness of the catalog is crucial (e.g., Wiemer 2001; Wiemer and Wyss 2000, 2002; Woessner and Wiemer 2005; Schorlemmer and Woessner 2008; Nanjo et al.

2010a,b). In this paper, we included an applied example by using the resultant $M_p$ to compute the parameters of GR distributions. We present a comparison between the spatiotemporal changes in $b$ and the $M$6.6 SSE whose fault model is shown in Fig. 1a. Crustal deformation due to this SSE, which grew over time during 2017-2018, caused stress perturbation in nearby regions. The overriding continental plate slipped with respect to the subducting Philippine Sea plate during the occurrence of the SSE, causing stress to accumulate in a down-dip volume of the Philippine Sea plate, while stress relaxed in an up-dip volume of the same plate. It remains uncertain whether the changes in $b$ in space and time, reflecting the state of stress (Scholz 1968, 2015; Smith 1981; Schorlemmer et al. 2005; Narteau et al. 2009), is applicable to assist image stress transferred by the slow slip of an SSE. The $M$6.6 SSE allows us to address this idea.

In creating Fig. 8a, which shows a time-series of $b$ values, the regions up-dip and down-dip of the SSE (blue and red regions in Fig. 1a), we used earthquakes in a depth range from the plate boundary to a depth of 25 [km] below the plate boundary (Supplementary Fig. S4). We then employed a moving window approach, whereby the window covers 80 earthquakes (this window is roughly one year). The MAXC method (Wiemer and Wyss 2000; Wiemer 2001) was applied to 80 earthquakes covered by the window for computing $M_{c(MAXC)}$. Then, for computing $b$, the maximum-likelihood method was applied to earthquakes with $M \geq M_{c(MAXC)} + M_{corr}$ (Woessner and Wiemer 2005). From our study (Fig. 6a-c), the maximum of $M_p$ values in the source area of the $M$6.6 SSE was ~2, irrespective of time in 2015-2019. We used this as prior information. We decided to select $M_{corr}$=0.5 in order for the $b$ value to be computed based on earthquakes with $M$~2 or higher. Fig. 8b confirms that $M_{c(MAXC)} + M_{corr}$ was around 2 throughout the entire investigation period. Note that a recommended value $M_{corr}$=0.2 (Woessner and Wiemer 2005) was not used for our case. Fig. 8a, which shows a time-series of $b$ values computed using samples with $M \geq M_{c(MAXC)} + M_{corr}$ with $M_{corr}$=0.5, is a product that considers $M_p$.

Although the MAXC is a method that has been used for a long time, some mathematical approaches were recently developed for calculating the magnitude of completeness, such as the use of the Lilliefors test (Herrmann et al. 2022). Further research requires to compare this method with our result to see if the 0.5 correction ($M_{corr}$=0.5) that was introduced is exaggerated and conservative.

The $b$ values for the up-dip and down-dip regions of the SSE (red and blue regions in Fig. 1a) were mostly around 1.5 before the start of the SSE, after which the $b$ values for the former region (thick red curve in Fig. 8a) showed an increase over time, to values around 2. In contrast, for the latter region, $b$ values showed a decrease over time, to values around 1 (thick blue curve in Fig. 8a). The difference in $b$ between these regions was significant on 2017.5 [decimal year] or later, taking the error bars of $b$ values into consideration.

The result is not induced by sampling bias (thin blue and red curves in Fig. 8a,b). We changed the sampling criterion slightly: sampling earthquakes at a depth range from 5 [km] above the plate boundary to a depth of 25 [km] below the plate boundary, in order to take a usual depth error (5 [km]) into consideration for earthquakes near the plate boundary (Supplementary Fig. S5). We applied the same plotting procedure as before. The data for both sampling cases (thick and thin curves in Fig. 8a,b) show the decreasing and increasing trends in $b$ during the SSE for the down-dip side (blue curves) and the up-dip side (red curves), respectively.

Moreover, the same analysis was conducted for a longer window covering 100 earthquakes (Supplementary Fig. S6), resulting in a similar result as that in Fig. 8a,b. Fig. 8c, which indicates the frequency-magnitude distribution of earthquakes after the start of SSE from the down-dip volume (blue) and the up-dip volume (red), shows a significant difference in $b$ between them. Our overall results support that the changes in $b$ reflect the changes in stress state caused by crustal deformation due to the SSE.

In Supplementary Fig. S7, the time-series of $b$ values for using the catalog that includes earthquakes in both of the red and blue regions shows similar behavior to that for the blue region (blue curves), but different behavior from that for the red region (red curves). This result demonstrates how specific the separation of the region of interest into the two regions (up- and down-dip regions of the SSE) was and confirms that the $b$-value variation (red versus blue curves shown in Fig. 8) was noticeable.

The same analysis as Fig. 8 was performed for using the JMA catalog. Supplementary Fig. S8 shows similar (but insignificant) outcome to Fig. 8. In creating Supplementary Fig. S8, the period from Apr. 1, 2016 to Oct. 1, 2020 was considered. For this period, homogeneous monitoring of the seismic networks used to create the JMA catalog was secured (https://www.data.jma.go.jp/eqev/data/bulletin/catalog/notes_j.html).

## 5. Discussion

The ability of DONET to detect an earthquake was evaluated in this study. We used a currently available catalog that includes earthquakes with $M$=-0.4~5.9 from Oct. 1, 2015 to Mar. 31, 2019 in and around the authoritative region of DONET (Fig. 1). From this catalog, earthquakes along the Philippine Sea plate were selected. The PMC method was applied to these earthquakes, and adjusted to the evaluation of DONET's performance. The PMC method is not based on selecting earthquakes over time and space, and $M_p$ is not the completeness of a volume from which catalog is selected. Instead, the PMC method is based on a station's characteristic, namely $P_D$ (Fig. 3b), and $M_p$ is the completeness defined at any location and time point in a given study region and investigation period. We considered the authoritative region of DONET (black rectangle in Fig. 1) as the study region, and used 0.05° × 0.05° maps as location points where $M_p$ was computed. Time points when $M_p$ was computed were selected: Jan. 1 for 2016, 2018, and 2019 (Fig. 6).

Regions of high probabilities ($P_E \geq 0.9$) in Fig. 5a that DONET detected an earthquake of $M=1$ for Jan. 1, 2019 covered almost all areas of DONET. $P_E \geq 0.9$ for $M=2$ (Fig. 5c) covered almost the entire authoritative region. To define $M_p$, we considered $P_E=0.999999$ (=1-$Q$ with $Q=10^{-6}$). For Jan. 1, 2019 (Fig. 6a), $M_p$ was <2 for the DONET1 and DONET2 areas while $M_p$ was >2 in between. $M_p$ was <1 in areas near Nodes D and E of DONET2.

The probability $P_E=0.999999$ was based on the assumption that earthquakes are independent, identically distributed events (e.g., Schorlemmer and Woessner, 2008; Nanjo et al., 2010b). In reality, however, there are many confounding factors such as depth, focal mechanism, aftershock sequences, and other spatiotemporal changes, that can significantly modify the probability of being able to detect an earthquake. Moreover, external causes such as meteorological and oceanographical events can also strongly affect the complementary probability $Q$. Thus, even if contour lines of $P_E=0.999999$ convey useful information, to avoid any misunderstanding, it is better to keep in mind that such factors and causes were not considered in this study. Only one issue among them was addressed in this study, namely external causes that were taken into account, using the network failure of DONET.

First, we evaluated the history of DONET in Oct. 2015-Mar. 2019. By computing the operational times of all stations for the investigation period (Fig. 2), we found the number of operational stations: 49, 46, 46, and 42 for Jan. 1 of 2016, 2017, 2018, and 2019, respectively. This decrease in number was mainly due to a malfunction of some individual stations and nodes (hubs that connect stations aggregated around nodes), but they were not repaired during the investigation period. The temporal evolution of $M_p$ (Fig. 6a-c) highlighted the pronounced effect on $M_p$ if those stations and nodes were not repaired.

We assumed that operation of stations connecting Nodes A and E of DONET1, which were interrupted since 2018 and 2016, respectively (Figs. 2a and 6a-c), had restarted. The area that covers the extent of these nodes corresponds to a deep part of the rupture region of the 1944 Showa Tonankai earthquake, and is associated with a transition between strong and weak plate-coupling zones. The transition zone may be the area at which initiation of future megathrust rupture is likely. Restarting the operation allowed us to intensively monitor seismicity around one of the target areas for future Nankai Trough events.

Second, we used PMC as a scenario computation tool to infer network performance for the virtual removal of stations. Two cases that correspond to a failure of Node E of DONET2 and a failure of another Node D next to the failed Node E showed max $\Delta M_p \sim 1.0$ (Fig. 7c) and 1.2 (Fig. 7d). Given that the earthquake detection capacity at sites near Nodes D and E was the highest in the study region (Fig. 6a) and that rupture initiation of the previous Nankai Trough events (the Showa Tonankai/Nankai earthquakes in 1940's) occurred at sites near Nodes D and E, failure of Node E or both Nodes D and E would be the most pronounced among nodes' failures. Repairing these nodes or replacing them with new ones if they malfunctioned, would be desired.

Even with the failure in operation of one or two nodes, no significant increase in $M_p$ (no significant lowering of earthquake detection capacity) occurs in surroundings of the areas that covered the extent of the failed nodes (Fig. 7c,d). This indicates that the strategy of aggregated configuration of stations around the respective nodes was successful, implying that this configuration minimizes the impact of node malfunction.

An example of how to use $M_p$ information as prior knowledge to seismicity-related studies was demonstrated (Fig. 8). Using the $b$ values, and taking $M_p$ into consideration, the changes in $b$ were found to be strongly correlated to the $M$6-class SSE. This is interpreted as an indication of the $b$ value as a proxy that can help to image stress heterogeneity when there is an SSE.

The reader may be curious about the comparison of the $b$ values obtained by DONET with those obtained by only onshore seismic networks. Nanjo and Yoshida (2018) used the JMA catalog to record earthquakes observed by only onshore seismic networks, and showed $b$ values along the Nankai Trough, including the current study region. There is an apparent difference in the absolute $b$ values between the two studies: the $b$ values in and around the SSE source area were 1~2.5 (Fig. 8) but they were $b$=0.9~1.2 in the study by Nanjo and Yoshida (2018). Similar outcome to Fig. 8 was obtained by using the JMA catalog (Supplementary Fig. S8) but this outcome was insignificant. Majority of the resultant $b$ values was 1~1.5.

We discuss the reason for these differences based on the comparison between $M$ and $M_{\mathrm{JMA}}$ in Fig. 3c for the offshore region, which includes the SSE source area. Since the tendency of $M>M_{\mathrm{JMA}}$ was more pronounced with decreasing $M$ (Fig. 3c), the data in the GR distribution for DONET were systematically shifted toward larger magnitudes, and this shift was larger for small $M$ than for large $M$. The shift in magnitude increased the slope of the GR distribution, suggesting this as a reason for the larger $b$ value for DONET than for JMA.

The main factor impacting the shift in magnitude is the difference between the equation to define magnitude employed by JAMSTEC (equation provided by Watanabe 1971) and that by JMA (equation proposed by Funasaki and Earthquake Prediction Information Division 2004). We think that factors beyond the magnitude definition are complex and influence $b$. One possible factor is site condition of the oceanfloor sediment where the DONET stations were installed. Due to soft oceanfloor sediments, site amplification during shaking is commonly recognized in observed motion records (Nakamura et al. 2013). Using the ground motion recorded by DONET, Watanabe's equation, which was not formulated for oceanfloor network systems such as DONET, very likely gave an overestimation of magnitude, resulting in high $b$ values. Another possible factor is increased uncertainties in magnitude determination and event location for the JMA catalog due to distant monitoring of offshore events from onshore networks. This was likely the cause of insignificant outcome shown in Supplementary Fig. S8. Further possible factors include difference in network operating procedures between JAMSTEC and JMA, and interaction of one or more of the above-described factors.

The $b$ values for the region's updip and downdip of the SSE are suggested by our results to behave differently during the SSE event. However, we noted that there was little explanation for the physical reasons for this behavior in this paper. It is of interest to point out Schurr et al. (2020) who used the Coulomb stress modeling and discussed behavior of up- and downdip of the 2014 $M$8.1 Iquique, Chile, earthquake. Two weeks before the mainshock, a series of foreshocks first broke the upper plate then the downdip rim of the asperity. This seismicity formed a ring around the slip patch (asperity) that later ruptured in the mainshock. Most features of the spatiotemporal seismicity pattern can be explained by a mechanical model that used the Coulomb stress in which a single asperity is stressed by relative plate motion.

Previous studies as well as our study related $b$ values to stress variations. Our result (Fig. 8) is an interesting observation, but further discussion on temporal variations on $b$ values and SSE/seismicity elsewhere might be needed. A specific example includes the 2014 Iquique earthquake sequence (Schurr et al. 2014). Leading up to this earthquake, the $b$ values of the foreshocks gradually decreased during the several years before the earthquake, reversing its trend a few days before the Iquique earthquake. Schurr et al. (2014) inferred that gradual unlocking (SSE) of the plate interface accentuated by the foreshock activity was instrumental in causing final failure. Similar decreases in $b$ values before large megathrust earthquakes have been documented, in particular for the $M$9 Tohoku, Japan, event (Nanjo et al. 2012).

The focus of our future research is two-fold:

1.  The reader may note that it is not straightforward for the current study to assess the robustness of the temporal change in the $b$ value associated with SSE because the data used for the evaluation is only three years (Fig. 8). JAMSTEC is updating a catalog to include recent events into it. A longer duration of the $b$ value timeseries is required for the robustness assessment.

2.  Our future work will be directed at showing spatiotemporal variability of the completeness magnitude and the monitoring capability in a real-time fashion through a website. That is because the Nankai Trough is one of the areas where seismicity should be monitored intensively.

## 6. Conclusion

1.  In this paper, we focused on assessing the earthquake detection capability of the DONET. We applied the methodology in the Nankai Trough, a region monitored for potential megathrust earthquakes. We used the earthquake catalog collected by the JAMSTEC from Oct. 2015 to Mar. 2019.

2.  In the first part of the paper, we evaluated the impact of network component malfunctions. We explained the completeness magnitude, the threshold above which all earthquakes are considered detected by a seismic network. We described the methodology applied in detail in the paper.

3.  In the second part, we observed that in the DONET area, the spatial and temporal variability in completeness magnitude ranged from values below 1 in densely populated station areas to values above 2 at the periphery and outside the DONET area.

4.  We used simulation computations for cases involving malfunctions of densely distributed stations. We

concluded that completeness estimates near malfunctioning stations were approximately 1 magnitude larger. We suggested that repairing or replacing malfunctioning stations would be desirable as it significantly affects earthquake monitoring.

5. In the paper's concluding section, we presented details on leveraging completeness magnitude data as prior information to calculate the $b$ value in the GR distribution. Our findings propose that the $b$-value can be a surrogate for visualizing stress heterogeneity during a $M$6-class slow slip event along the Nankai Trough plate boundary.

**Acknowledgments** The authors thank Editor (Klaus-G. Hinzen) and three anonymous referees for their valuable comments, JAMSTEC and JMA for the earthquake catalogs, and NIED for help creating Supplementary Figs S1-S3.

**Data** Waveform data recorded by the DONET stations are available at https://doi.org/10.17598/nied.0008. The JMA earthquake catalog used in this study was obtained from http://www.data.jma.go.jp/svd/eqev/data/bulletin/index_e.html. The seismicity analysis software package ZMAP, used for Fig. 4 and 8 and Supplementary Figs. S4-S8, was obtained from http://www.seismo.ethz.ch/en/research-and-teaching/products-software/software/ZMAP. The Generic Mapping Tools (GMT), used for Figs. 1a, 2c, and 5-7, are an open-source collection (https://www.generic-mapping-tools.org) (Wessel et al. 2013). Codes of the PMC method are available from the corresponding author upon reasonable request.

**Author Contribution** KZN, YY and KA designed the study. KZN conceived the method of analysis, created figures, and wrote the paper. Preparation of earthquake data used in this study was carried out by YY, HH, SY, and NT. All authors developed the manuscript and approved the final manuscript.

**Funding** This study was partially supported by Chubu Electric Power's research based on selected proposals (K.Z.N., Y.Y., K.A., N.T.), the Ministry of Education, Culture, Sports, Science and Technology (MEXT) of Japan, under its Earthquake and Volcano Hazards Observation and Research Program (K.Z.N.) and under the STAR-E (Seismology TowArd Research innovation with data of Earthquake) Program Grant Number JPJ010217 (K.Z.N.), and JSPS KAKENHI Grant Number 22K03766 (Y.Y.).

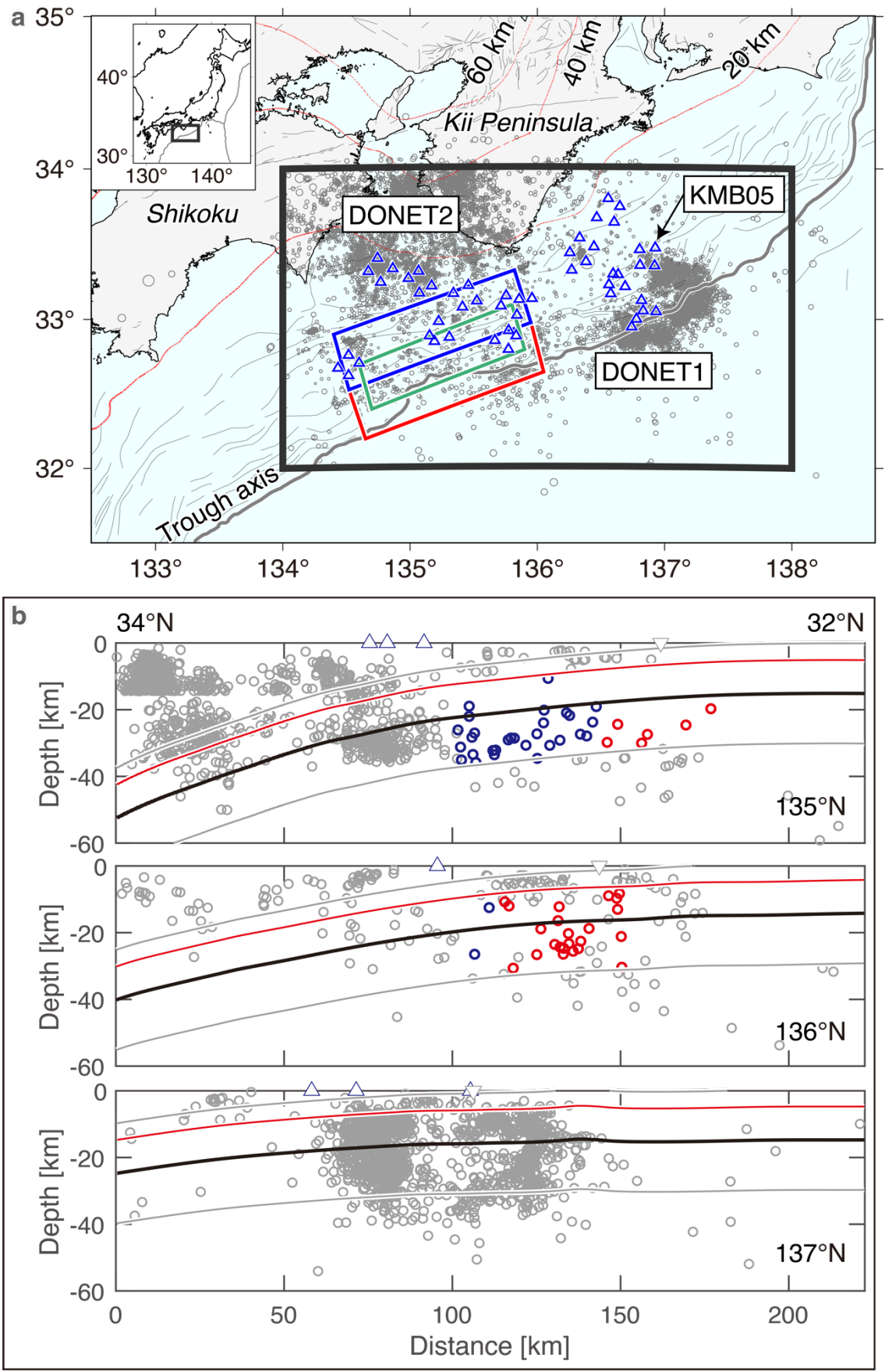

**Fig. 1** Distribution of earthquakes with a depth of 0-60 [km] in 2015-2019. **a**, Map view showing earthquakes with *M*≥0, where the radius of a circle is proportional to magnitude. Triangles show stations

within DONET. Stations within subnetworks DONET1 and DONET2 are separated at 136°E. Rectangular region in black shows the authoritative region of DONET. Station KMB05 is highlighted. Green rectangle indicates the 2017-2018 shallow SSE model (Yokota and Ishikawa 2020). Blue and red rectangles show the regions in which earthquakes were used to create Fig. 8. Thick grey curve indicates the trough axis. Red curves indicate depth contours (20, 40, and 60 [km]) of the plate boundary (the upper surface of the subducting Philippine Sea plate: Nakanishi et al., 2018). Thin grey segments show active faults. Inset shows the authoritative region of DONET (black rectangle). **b**, Cross-sectional views showing hypocentral distribution of earthquakes given in each north-south zone with a width of 20 [km] at 135, 136, and 137°E. Blue and red circles indicate earthquakes in the blue and red rectangles shown in **a** and used to create Fig. 8. Red curve represents the upper surface of the subducting Philippine Sea plate. Black curve is drawn at a depth of 10 [km] below this surface. Values of $P_E(M, x)$ in Fig. 5 and $M_p(x)$ in Fig. 6, and $M_{pv}$ and $\Delta M_{pv}$ in Fig. 7 were computed at this depth. Grey curves are drawn at depths of 5 [km] above the surface and 25 [km] below the surface. Upward-pointing triangles indicate stations within DONET. Downward-pointing triangles indicate the trough axis.

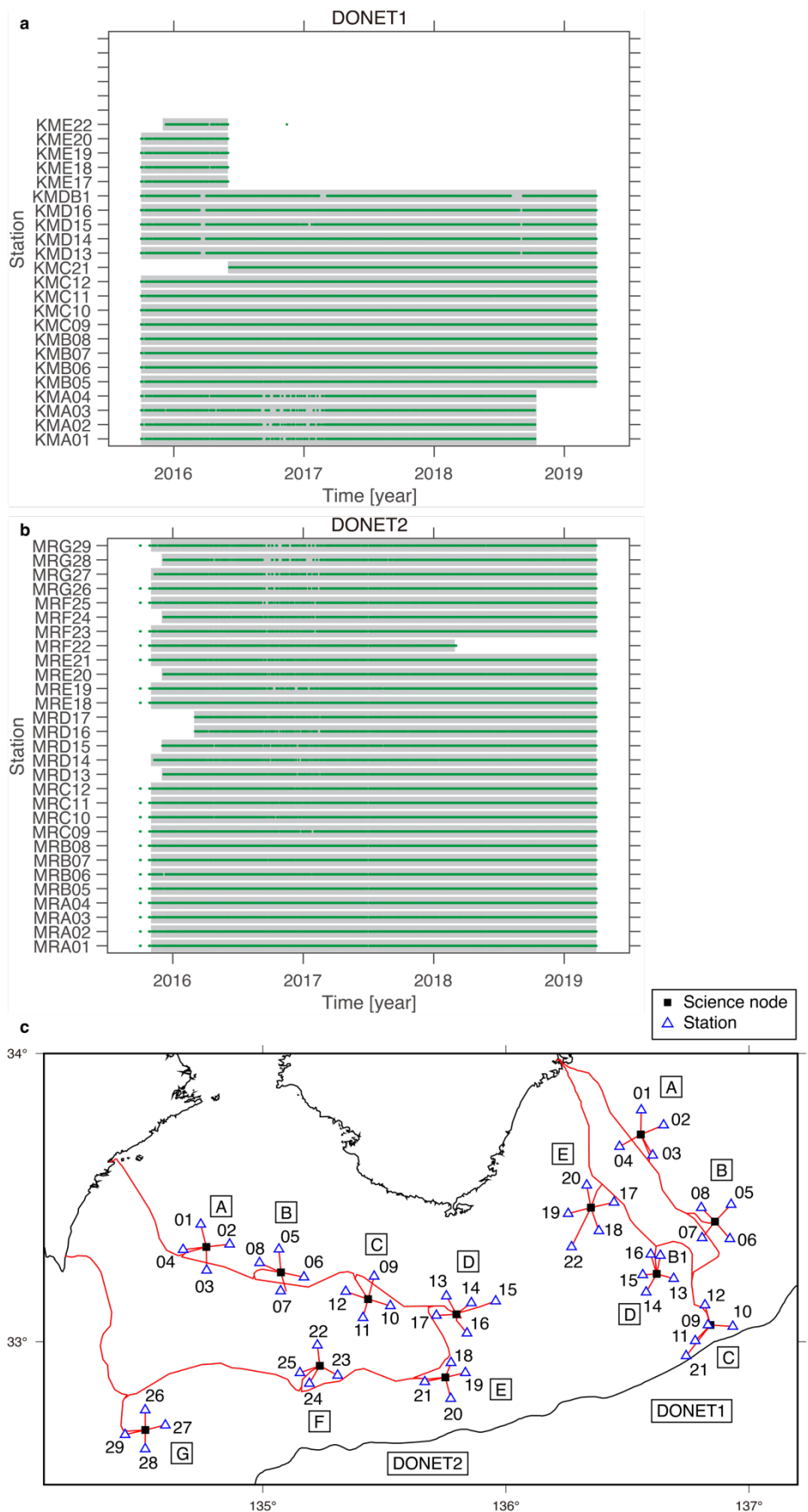

**Fig. 2** Station information. **a**,**b**, Operational times for stations within DONET1 and DONET2. Station names are indicated on the left axis. If the first two characters of the station name indicate "KM" or "MR", the station is within DONET1 or DONET2, respectively. The third letter (A, B, … E for DONET1 and A, B, … G for DONET2) indicates the station connecting the science node identified by the corresponding letter. The last two characters (number or letter of the alphabet "B") indicate station ID. KMDB1 is the borehole station (Kopf et al. 2011). For each station, the green dots indicate the times when the station was used to record earthquakes, and the grey bars indicate its estimated operational times. **c**, Map showing stations (triangles), science nodes (square), and backbone cables (lines). The letters in squares indicate the letters of the alphabet that identify each node. Two characters (number or letter of the alphabet "B") next to triangles indicate station ID.

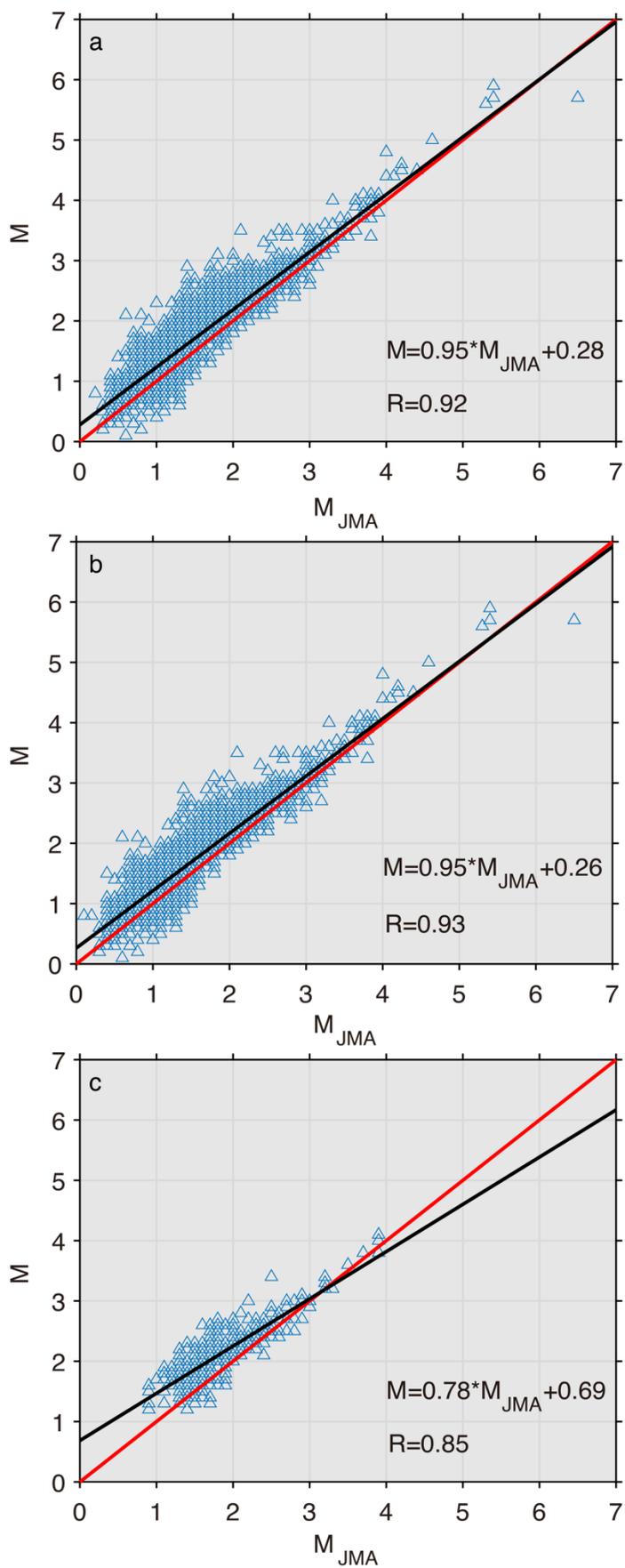

**Fig. 3** Difference in magnitude **a**, The DONET magnitude (*M*) is plotted as a function of the JMA magnitude

($M_{JMA}$) for the authoritative region of DONET (rectangular region in black in Fig. 1a: 134-138°E, 32-34°N,

depth of 0-60 [km]). Black and red lines indicate the least-square regression line and $M=M_{\text{JMA}}$, respectively. The equation with the correlation coefficient ($R$) for the least-square regression line is shown. **b**, Same as **a** for the nearshore region (134-138°E, 33-34°N, depth of 0-60 [km]). **c**, Same as **a** for the offshore region (134-138°E, 32-33°N, depth of 0-60 [km]).

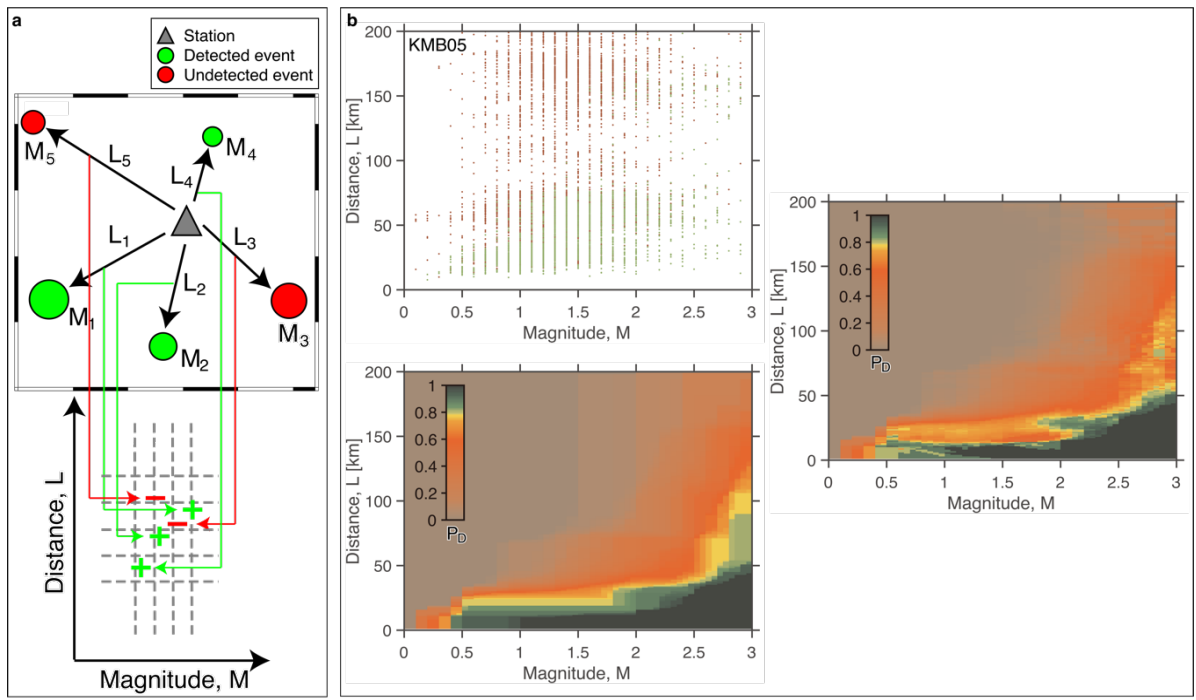

**Fig. 4** Station characteristic. **a**, Illustration showing the procedure to create a distribution of detected and undetected events (a modified version of Fig. 1 of Schorlemmer and Woessner 2008). Top panel: spatial map of five events (event 1, event 2, … event 5) detected by DONET, where the magnitude of an event $i$ is $M_i$ and the distance of event $i$ from the station is $L_i$ with event number $i$ = 1, 2, … 5. Bottom panel: graph of $L$ as a function of $M$. If the station was used to detect an event, the data triplet of this event is called a "plus triplet", colored in green in the map (top panel) and plotted in the graph by a green plus symbol (bottom panel). If the station was not used, the data triplet is called a "minus triplet" (red, minus symbol). **b**, Detection characteristic of station KMB05, highlighted in Fig. 1a. (Top) Distribution of 2100 unpicked (red) events and 3611 picked (green) events for the station. (Middle) Distribution of detection probabilities over the magnitude and distance of station KMB05 derived from raw data triplets. (Bottom) Smoothed distribution of detection probabilities.

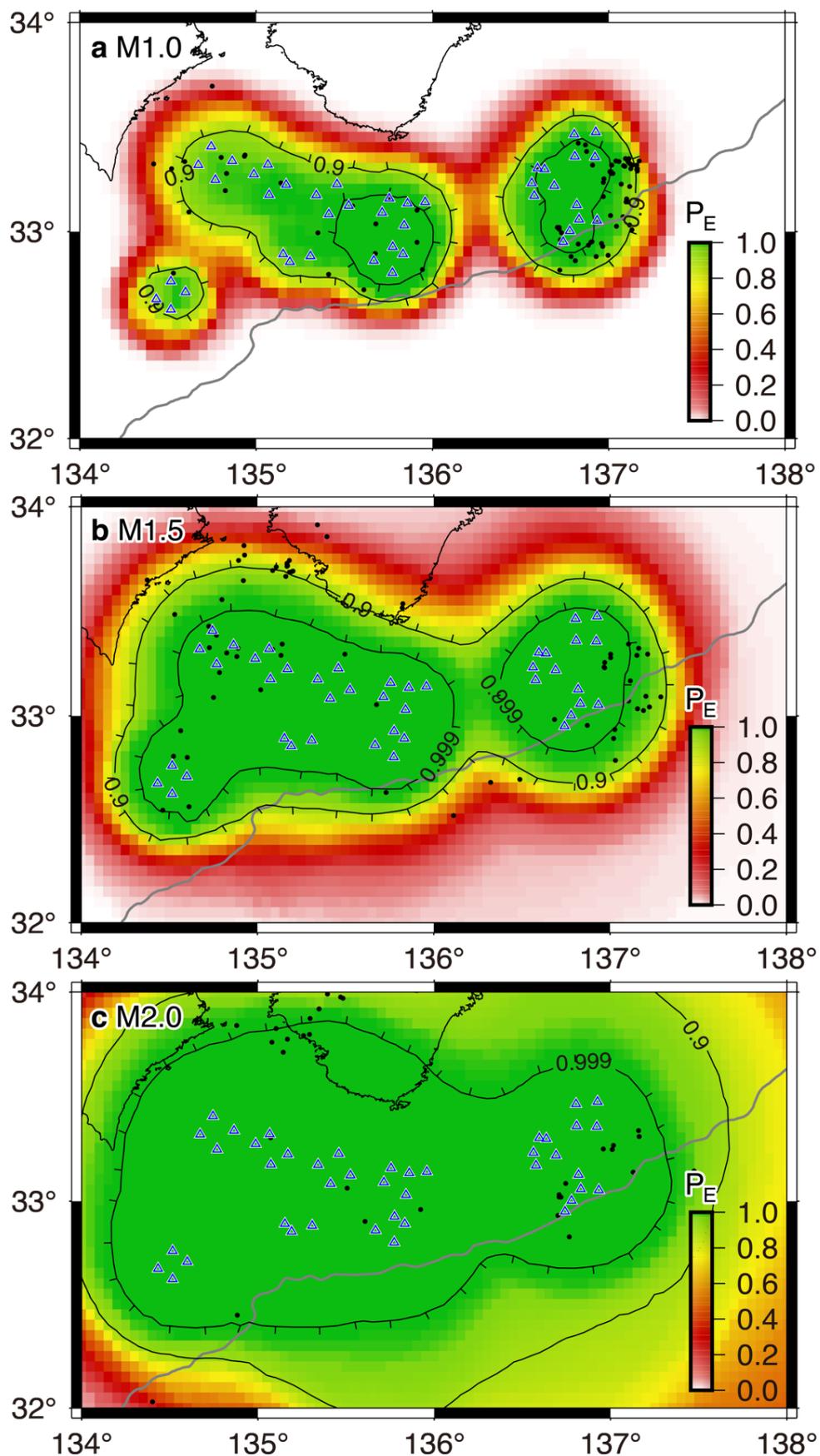

**Fig. 5** Maps of $P_E$. The values were computed at a depth of 10 [km] below the upper surface of the Philippine

Sea plate (black curve drawn in Fig. 1**b**). **a-c**, $P_E$ for $M$=1.0, 1.5, and 2.0 on Jan. 1, 2019. Dots show events from the period Oct. 16, 2018-Mar. 31, 2019. See the caption of Fig. 1 for an explanation of thin red curves, blue triangles, and thick grey curve.

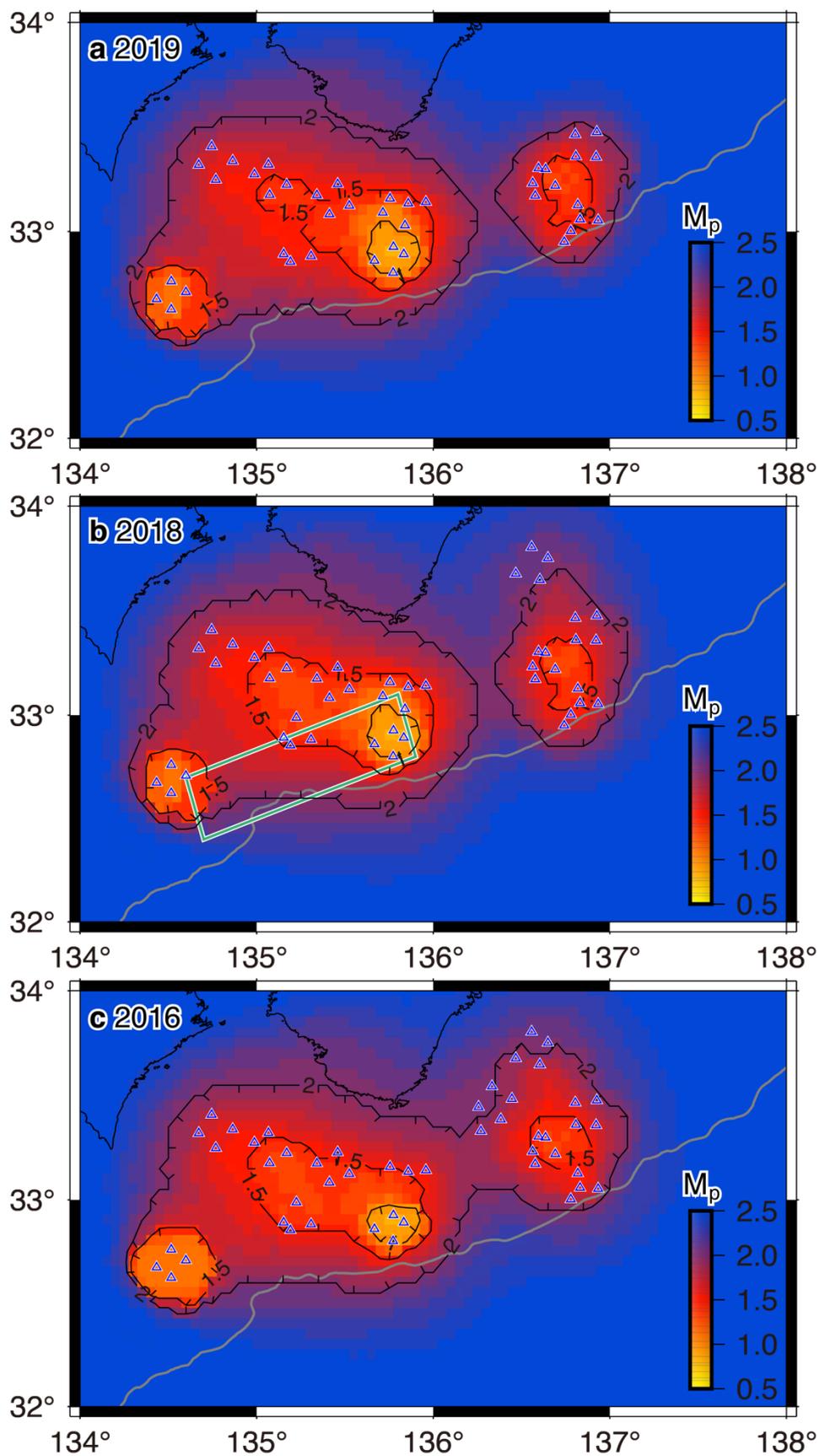

**Fig. 6** Maps of $M_p$. The values were computed at a depth of 10 [km] below the upper surface of the Philippine Sea plate (black curve drawn in Fig. 1**b**). **a-c**, $M_p$ at Jan. 1 of the respective years as indicated in the frames (2019, 2018, 2016). Green rectangle in **b** indicates the 2017-2018 shallow SSE model (Yokota and Ishikawa 2020). The panels imply that outside the DONET, $M_p$ is constant with 2.5 (blue), but $M_p$ further increases with increasing distance from the network. See the caption of Fig. 1 for an explanation of blue triangles and thick grey curve.

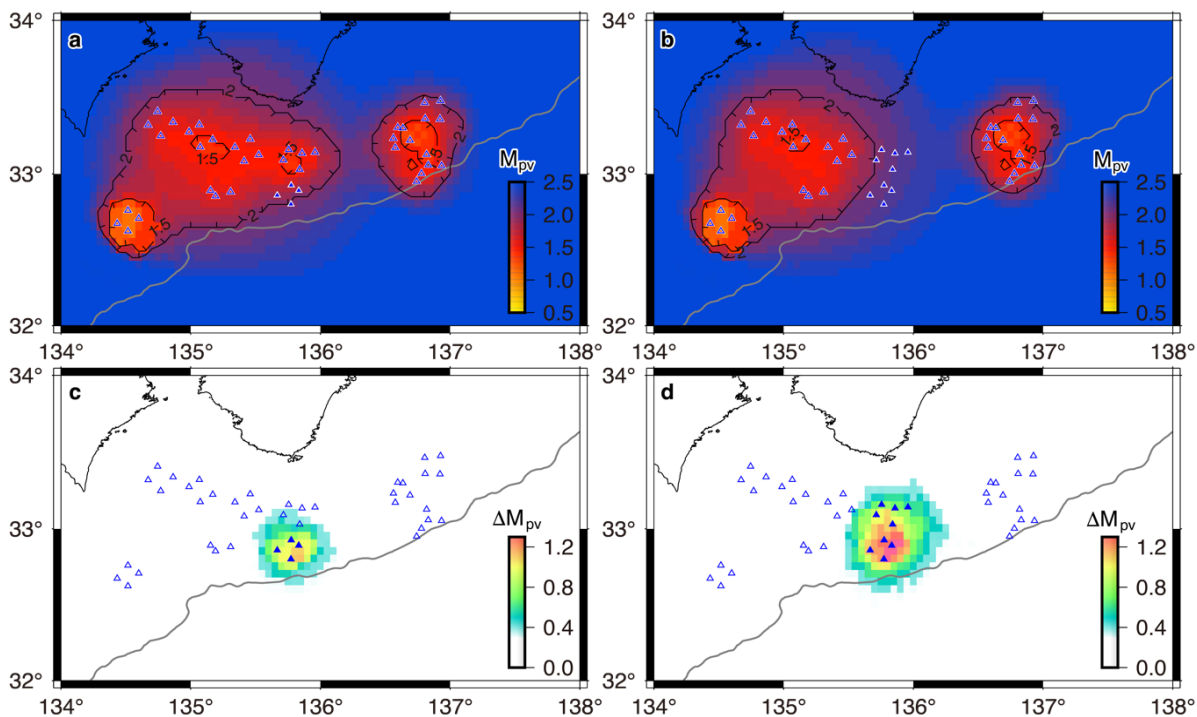

**Fig. 7** Scenario computations by removing the nodes from DONET to examine their effect on network detection capacities. Nodes were virtually removed from DONET in the operation on Jan. 1, 2019 (Fig. 6**a**). **a**, Same as Fig. 6**a** for removing Node E of DONET2 (filled triangle). Operating stations are indicated by open triangles. **b**, Same as **a** for removing another Node D of DONET2. **c**, Color code indicates the difference in completeness magnitude ($\Delta M_p$) between the virtual case ($M_{pv}$) in **a** and the real case ($M_p$) in Fig. 6**a**. Locations of virtually removed stations are indicated by filled triangles. **d**, Same as **c** for $\Delta M_p$ between $M_{pv}$ in **b** and $M_p$ in Fig. 6**a**. See the caption of Fig. 1 for an explanation of the thick grey curve.

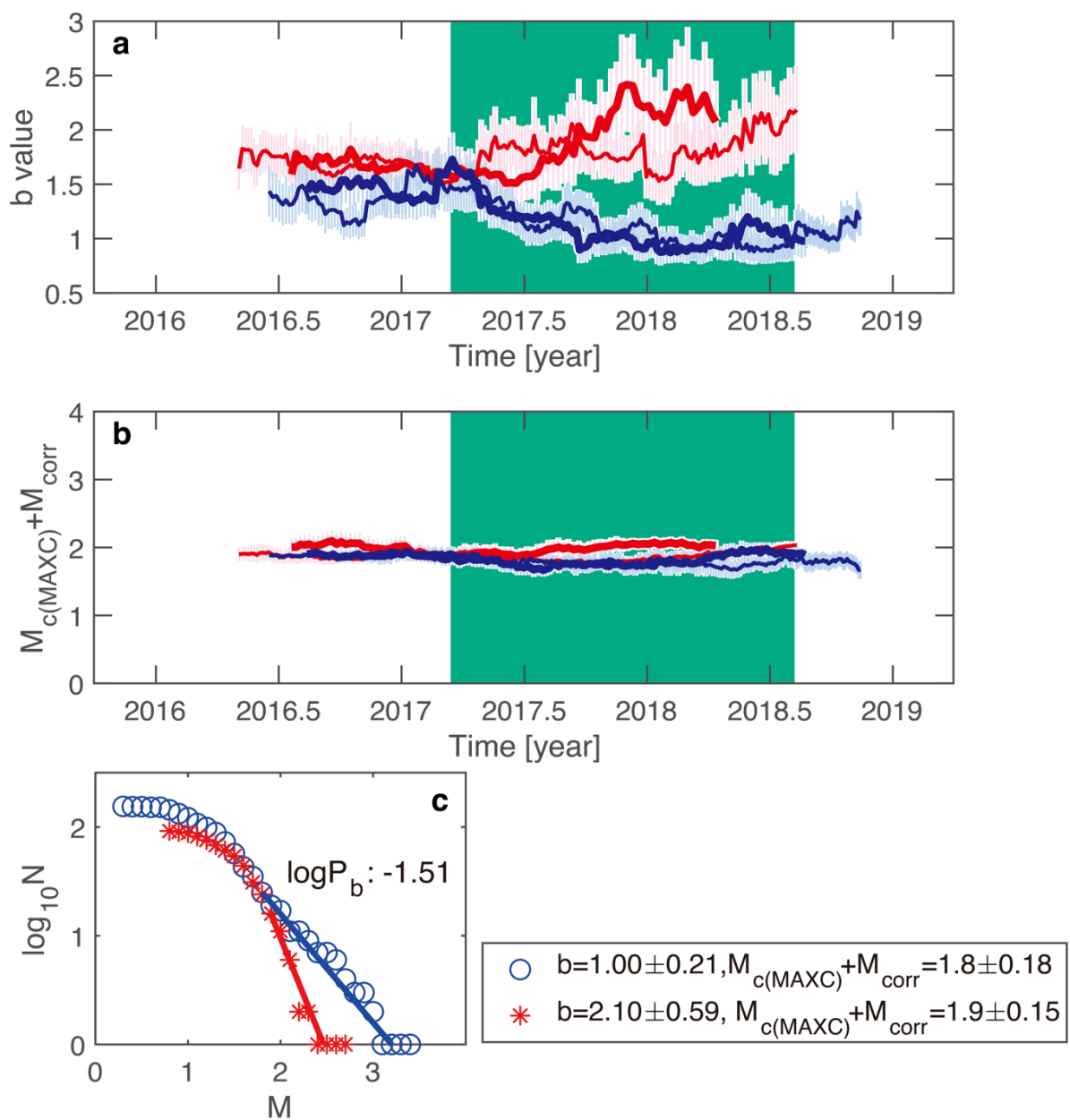

**Fig. 8** Plot of *b* as a function of time. **a**, Thick red curve: *b*-values calculated based on seismicity in the red rectangular region in Fig. 1**a** in the depth range from the upper surface of the Philippine Sea plate to 25 [km] below this surface. A moving window approach was used, whereby the window covered 80 events. Each *b* value was computed for events with $M \geq M_{c(MAXC)} + M_{corr}$, and plotted at the mean of the occurrence times of earthquakes covered by the window, where $M_{corr}=0.5$ was chosen (Supplementary Fig. S4). Thin red curve: same as the thick red curve for the depth range from 5 [km] above the surface to 25 [km] below the surface (Supplementary Fig. S5). Thick and thin blue curves: same as the thick and thin red curves for seismicity in the blue rectangular region in Fig. 1**a** (Supplementary Figs. S4 and S5). Green region indicates the period of the *M*6.6 SSE (Yokota and Ishikawa 2020). Location of its source fault model is indicated by a green

rectangular in Fig. 1**a**. **b**, Same as the top panel for $M_{c(MAXC)}+M_{corr}$. The error bars in **a** and **b** represent one standard deviation of a bootstrap sampling distribution (Schorlemmer et al. 2003). **c**, Frequency-magnitude distribution of earthquakes. Data indicated by red asterisks and blue circles are the same used for the thick red and blue curves, respectively, in the time-series in **a** and **b**, except that we used earthquakes after the start of SSE (after 2017.2 [decimal year]). GR fitting is given by the lines, where the procedure to compute $b$ and $M_{c(MAXC)}+M_{corr}$ is the same as that used in **a** and **b**. $\log P_b = 1.51$ indicates a significant difference in $b$ (Utsu 1992, 1999; Schorlemmer et al. 2004; Nanjo and Yoshida 2017; Nanjo et al. 2019; Nanjo 2020b). This observation was further supported by applying bootstrapping errors (Schorlemmer et al. 2003; Nanjo and Yoshida 2017; Nanjo et al. 2019; Nanjo 2020b): the range of $b$ values for blue and red data do not overlap with each other.